# Strain-induced orbital energy shift in antiferromagnetic $RuO_2$ revealed by resonant elastic x-ray scattering

Benjamin Gregory[1,2], Jörg Strempfer[3], Daniel Weinstock[2], Jacob Ruf[4], Yifei Sun[2], Hari Nair[2], Nathaniel J. Schreiber[2], Darrell G. Schlom[2,5,6], Kyle M. Shen[1,5], and Andrej Singer[2,*]

[1]*Laboratory of Atomic and Solid State Physics, Department of Physics, Cornell University, Ithaca, NY 14853, USA*
[2]*Department of Materials Science and Engineering, Cornell University, Ithaca, NY 14853, USA*
[3]*Advanced Photon Source, Argonne National Laboratory, Lemont, IL 60439, USA*
[4]*Max-Planck Institute for Chemical Physics of Solids, Nöthnitzer Straße 40, 01187 Dresden, Germany*
[5]*Kavli Institute at Cornell for Nanoscale Science, Cornell University, Ithaca, NY 14853, USA*
[6]*Leibniz-Institut für Kristallzüchtung, Max-Born-Straße 2, 12489 Berlin, Germany*

**In its ground state, $RuO_2$ was long thought to be an ordinary metallic paramagnet. Recent neutron and x-ray diffraction revealed that bulk $RuO_2$ is an antiferromagnet (AFM) with $T_N$ above 300 K. Furthermore, epitaxial strain induces novel superconductivity in thin films of $RuO_2$ below 2 K. Here, we present a resonant elastic x-ray scattering (REXS) study at the Ru $L_2$ edge of the strained $RuO_2$ films exhibiting the strain-induced superconductivity. We observe an azimuthal modulation of the 100 Bragg peak consistent with canted AFM found in bulk. Most notably, in the strained films displaying novel superconductivity, we observe a ~1 eV shift of the Ru $e_g$ orbitals to a higher energy. The energy shift is smaller in thicker, relaxed films and films with a different strain direction. Our results provide further evidence of the utility of epitaxial strain as a tuning parameter in complex oxides.**

The ruthenium-based oxides host a rich set of physical phenomena, including unconventional superconductivity in $Sr_2RuO_4$ [1], a metamagnetic ground state in $Sr_3Ru_2O_7$ [2], insulating, antiferromagnetism in $Ca_2RuO_4$ [3] and $Ca_3Ru_2O_7$ [4], and both paramagnetic and ferromagnetic metallic states in $CaRuO_3$ and $SrRuO_3$, respectively [5]. Ruddlesden-Popper ruthenates undergo a variety of electronic, magnetic, and orbital ordering transitions, which are tunable with chemical doping, pressure, temperature, magnetic field, and epitaxial strain [6] . In the ruthenium-based superconductor $Sr_2RuO_4$, uniaxial pressure has been shown to increase $T_c$ [7] and epitaxial strain can alter the topology of its Fermi surface [8]. Uniaxial pressure has additionally been shown to induce a paramagnetic to ferromagnetic transition in $Sr_3Ru_2O_7$ [9] and epitaxial strain can enhance the existing magnetization in ferromagnetic $SrRuO_3$ [10,11].

A particularly striking example in this vein is the recent creation of a novel superconductor with epitaxial strain from non-superconducting $RuO_2$ [12,13]. Bulk $RuO_2$ has a rutile crystal structure (space group #136, $a$ = 4.492 Å, $c$ =3.106 Å) at room temperature (295 K). Ruf et al. [12] reported strain-induced superconductivity in $RuO_2$ films, synthesized via molecular-beam epitaxy (MBE) on isostructural $TiO_2$ substrates ($a$ = 4.594 Å, $c$ = 2.959 Å) of differing surface orientations: $TiO_2$ (110) and $TiO_2$ (101). The superconducting ground state below 2 K is only present in the (110)-oriented films while the (101)-oriented films remain metallic. On $TiO_2$ (101) the lattice mismatch of the sample and substrate imparts in-plane tensile strain of +0.04% along $[\bar{1}01]$ and +2.3% along [010]. In the (110)-oriented sample the strain is compressive (-4.7%) along [001] and tensile (+2.3%) along $[1\bar{1}0]$. The unit cells including strain directions are shown in Figs. 1(a) and 1(b).

In addition to its unexpected superconductivity, $RuO_2$ was long thought to be an ordinary, metallic paramagnet [14], but recent neutron diffraction results support an antiferromagnetic (AFM) ground state with spins aligned along the $c$-axis with a Néel temperature, $T_N$, greater than 300 K [15]. Further resolution of the magnetic ordering is experimentally accessible through resonant elastic x-ray scattering (REXS) [16,17]. Azimuthal analysis of REXS (rotating the sample around the scattering vector) at the Ru $L_2$ edge in bulk $RuO_2$ supports the existence of AFM ordering with moments largely along the $c$-axis but with canting toward the $a$-$b$ plane [18]. This canted AFM conclusion has been questioned in subsequent work [19], though non-scattering techniques based on antiferromagnetic spin Hall effect offer further support for an antiferromagnetic origin of this signal [20]. In this letter, we use REXS to study the effect of epitaxial strain on the antiferromagnetism in strained films that display novel, strain-stabilized superconductivity [12]. We provide further analysis of the strain-dependent phenomenology of these samples, using "sc110" and "ns101" as shorthand for the superconducting (110)-oriented sample (thickness 21.0 nm) and the non-superconducting, (101)-oriented sample (thickness 18.6 nm), respectively.

To investigate the effect of large epitaxial strain on the magnetic ground state of the $RuO_2$, we performed resonant magnetic x-ray scattering at the Ru $L_2$ edge (2.968 keV) at beam line 4-ID-D of the Advanced Photon Source at Argonne National Laboratory (see inset in Fig. 1(c) for the scattering geometry). On resonance, superlattice peaks or structurally forbidden peaks appear corresponding to charge, spin, or orbital ordering. The polarization and azimuth dependence of these forbidden reflections have already provided refinement of the ordered phases in ruthenates [18,21–25]. We tuned the x-ray scattering geometry to the structurally forbidden 100 Bragg peak, shown to be sensitive to the AFM [18]. Because we used a vertical scattering geometry without polarization analysis, we averaged the scattered intensity in the $\sigma - \pi'$ and $\sigma - \sigma'$ channels. Figure 1(c) shows the normalized, integrated 100 peak intensity as a function of temperature for both strain orientations, upon cooling and heating. In addition, we have redrawn the original neutron scattering data from 2017 in the same plot [15]. Among all of these measurements, we observe general agreement in the temperature dependence, with a small discrepancy between the ns101 heating and cooling data. As neutrons offer a direct probe of magnetic order, we interpret the agreement of the temperature dependence measured here with the neutron data in bulk as further confirmation of the antiferromagnetic origin of the forbidden peak as opposed to another anomalous scattering mechanism.

Figures 2(a) and 2(b) show a width comparison of the structurally-allowed 200 Bragg reflection and the structurally-forbidden magnetic 100 Bragg reflection orthogonal to the scattering vector measured while rocking the incident angle $\theta$ (see inset in Fig. 1(c)). To access the 100 and 200 peaks, we tilted both samples normal to 45° with respect to the scattering plane, shown in the inset to Fig. 1(c). Due to the differing sample orientations, the peak widths in ns101 and sc110 correspond to correlation lengths along the [010] and [001] directions. In both films, the magnetic peak is broader than the structural peak. Assuming minimal mosaicity as a result of epitaxial growth, we attribute the broadening to the finite coherence length of the probed region, suggesting the presence of multiple magnetic subdomains present in each coherent crystalline domain. We quantify this analysis by estimating the size of the crystalline coherent (charge) domains and the magnetic domains from the full width at half maximum (FWHM) of the Bragg peaks. These estimates provide a lower bound on the domain sizes, whose estimated size would grow if mosaicity were accounted for. In ns101, we estimate the structural and magnetic domains to be around 710 nm and 230 nm along the [010] direction, respectively. In sc110, we estimate the structural and magnetic domains to be around 2.6 $\mu$m and 460 nm along [001], respectively. We thus find that each crystalline coherent domain contains 3-5 antiferromagnetic domains on average. Additionally, we observe that both the 100 and 200 peaks are broader in the ns101 film and suggest two possible explanations. First, the ns101 film may have smaller crystalline domains simply due to natural variation in the quality of epitaxial growth across samples. Second, the difference in widths may reflect that the domains are elongated in one direction since these line scans represent sections along different directions in the reciprocal space.

To determine if the different strain states change the orientation of the magnetic moments, we studied the azimuthal dependence of the magnetic Bragg reflection by measuring the integral intensity of the 100 peak while rotating the specimen around the momentum transfer vector **Q** by the azimuth angle $\psi$ ($\psi = 0$ when [001] lies in the scattering plane). The scattering geometry (see Fig. 1(c)) severely restricted the accessible azimuthal range, allowing only ~25° of rotation before the sample eclipsed the incident or diffracted x-ray beam. When the sc110 sample is in the right scattering geometry to access 100, the *c*-axis lies in the scattering plane, but to achieve the same Bragg condition in ns101, the *c*-axis points orthogonal to the scattering plane. Thus, in our azimuthal measurements we are able to access the region around $\psi = 0°$ in sc110 and around $\psi = 90°$ in ns101. Due to the scattering geometry (specifically that $\psi$ is around 90° in ns101, see Fig. 2), the REXS intensity from ns101 is much lower than that of sc110, and it may be that the scattering is sensitive to a small, unknown hysteretic effect that only appears when magnetic scattering signal is weak. For example, a small change in the canting of the spins toward the *c*-axis upon heating or cooling would show a larger percentage change of integrated intensity when the intensity is close to zero. Nevertheless, Figs. 2(c) and 2(d) both display a modulation of the integrated intensity with rotating azimuth in epitaxial films comparable to that measured in bulk $RuO_2$ [18], a further indication of the superb film quality and the absence of twinning in our samples.

The simplest model of resonant magnetic scattering consists of a magnetic moment in spherical symmetry with fixed incoming and outgoing, linear polarization. The leading contribution to the intensity of such a scattering process is [26]

$$I \propto |\varepsilon \times \varepsilon' \cdot \hat{m}|^2. \tag{1}$$

We first assume that the magnetic moment lies along the *c*-axis [16,24] and consider the orthorhombic ($D_{2h}$) crystal field around the Ru atom. In this case without canting ($\hat{m} = [001]$), the scattering tensor reduces to the same form as that in spherical symmetry [16]. We first analyze the sc110 data. After rescaling the amplitude to 1.0 and including an offset to account for possible experimental misalignment, $\delta$, the expected azimuthal dependence in the $\sigma - \pi'$ channel is

$$I(\psi) = \cos^2(\psi - \delta),\tag{2}$$

and zero in the $\sigma - \sigma'$ channel. This fit is shown in Fig. 2(c) with an offset of $\delta$ =-4(3)° and deviates significantly from the data with a large mismatch in periodicity. Both the crystal field around the Ru ion and the canting of its spin can further modulate the azimuthal dependence by lowering the local symmetry of the scattering atom [16]. To improve the fit we included a general, second harmonic term (half the period), that adds a four-fold modulation capturing contributions to the scattering tensor from reduced point symmetry without reference to a specific model:

$$I(\psi) = \left[A_1 \cos(\psi - \delta) + A_2 \cos(2(\psi - \delta))\right]^2.\tag{3}$$

We find a best fit with $A_1$ = 0.20(4), $A_2$ = 0.80(4), and $\delta$ = - 4.0(2)°. This improved fit is shown in green in Fig 2(c). Because of the short range of data, we are unable to verify more complex models without a severe overfitting. Nonetheless, we conclude that the inclusion of a $2\psi$ harmonic contribution is essential to reproduce the sc110 data. Because in the ns101 film the maximum value of the sinusoid in Fig. 2(d) is inaccessible due to the experimental geometry, we cannot measure the periodicity of its azimuthal dependence directly. Fortunately, the position of its minimum near 90° remains informative, since the $A_2$ term necessary to fit the sc110 data achieves a maximum at this value of $\psi$ and therefore cannot dominate the fitting in ns101. Accordingly, we find equation (2) can sufficiently describe the ns101 data with an offset of $\delta$ = 1.3(3)° and if we include the $A_2$ term in Eq. (3), its fitted value is zero ($A_2$ = 0.1(2)). The best attempt to force the model to fit both datasets simultaneously is shown in Fig. S1 of the Supplementary Material, but the resulting parameters are unphysical. Even with the limited azimuthal range accessible, the data are sufficient to show that substantially different parameters in Eq. (3) are required to describe the different samples.

This difference in the relative contributions of the two terms in Eq. (3) arises from differing local symmetry of the scattering center in the two samples. The authors in [18] identify a four-fold contribution to the azimuthal scan in bulk $RuO_2$ with terms corresponding to nonzero components of the magnetization in the *a-b* plane i.e., off *c*-axis spin canting. Our data show that such a four-fold term is the dominant contribution necessary to describe sc110 and at most a small perturbation in the case of ns101. Assuming that spin canting is the mechanism modulating the intensity with azimuth, then our findings suggest that the high compressive strain in sc110 alters the magnetization direction relative to bulk and bulk-like ns101. More recently the conclusion that canting can explain the azimuthal dependence has been called into question [19]. Those authors propose that a chiral signature calculated from all scattering channels is a better indicator of long range antiferromagnetism. Since our experiment used photons with linear polarization $\sigma$ we can offer no support nor counterevidence for such claims.

To explore the electronic behavior of magnetic Ru ions in the crystal environment, we measured the photon energy dependence of the magnetic Bragg reflection intensity 100 in the vicinity the Ru $L_2$ edge. The averaged energy scans of both films, sc110 and ns101, are shown together in Fig. 3. Details of this averaging are shown in Fig. S2. The double peak shape of the resonance has been previously observed in REXS studies of Ru-based oxides, with the low energy peak corresponding to transitions from $2p_{1/2}$ core level to $4d$ $t_{2g}$ orbitals and the high energy peak to transitions into $4d$ $e_g$ orbitals [18,21,25]. The most prominent feature in these resonances is the 0.93 eV shift of the $e_g$ peak to the higher energy in the superconducting sc110 sample. Notably the $e_g$ peak only shifts in the sample whose strain induces superconductivity, whereas the resonance profile of differently-strained, non-superconducting ns101 continues to resemble bulk.

To investigate the strain dependence of this peak shift, we measured the resonance profile of an identical, (110)-oriented sample, but of greater thickness (48 nm) denoted "th110" shown in Fig. 3. The thicker sample cannot remain uniformly strained throughout its whole thickness, so the lattice partially relaxes with an intermediate, average lattice constant more similar to bulk [12]. We observe that the partial relaxation leads to a leftward shift of the $e_g$ peak, more in line with bulk and ns101. This control measurement confirms that strain is the dominant origin for the $e_g$ energy shift in sc110. We detect no shift in the position of the lower energy $t_{2g}$ peak between sc110 and ns101. Because the photon energy was not calibrated for the th110 measurement, we shifted the resonance profile of th110 to align the low energy peak with the other two samples in order to compare the shift of its high-energy peak.

In addition to energy shifts, the relative intensities within the double peak profile varies across differently strained samples. We consider the possibility that the variation in $e_g$ peak intensity and line shape result from self-absorption. We would expect th110 to exhibit a smaller $e_g$ peak than sc110 as the former is twice as thick as the latter. The $e_g$ peak of th110, however, is about 1.5 times taller than that of sc110. Furthermore, the tabulated attenuation lengths below and above the $L_2$ edge are approximately 1200 nm and 900 nm, respectively [27]. Further assuming the absorption length decreases to ~500 nm on resonance, even in our thickest sample (48 nm), the x-ray intensity would fall by at most 5%, far too small to explain the variation in the $e_g$ peak intensities we measure.

To explain the strain and orbital dependence of the energy scan peak shift, we consider the spatial orientation of the Ru $4d$ orbitals in the strained, octahedral crystal field (Fig. 4). In an octahedral crystal field, the $t_{2g}$ set couples more weakly to the lattice as the lobes of the $d$ orbitals point toward the faces of the oxygen octahedra leading to weaker $\pi$ bonding and subsequently reducing their sensitivity to lattice distortions [28]. The lobes of $e_g$ orbitals point toward the vertices of the octahedra and form antibonding sigma molecular orbitals with the oxygen $2p$ states. Thus, the $e_g$ set is more sensitive to changes in bond length with increasing atomic orbital overlap raising the energy of the molecular orbital. Figure 4(a) illustrates displacement of the oxygens toward the $d_{x^2-y^2}$ and $d_{z^2}$ orbitals under the strain applied in sc110. Though the oxygens are displaced by +2.3% away from the $d_{z^2}$ along the $[1\bar{1}0]$ direction, this relaxation is smaller than the predominant $c$-axis compression of -4.7%, with the net effect of raising the energy of the $e_g$ set. The $d_{x^2-y^2}$ and $d_{z^2}$ orbitals under +2.3% $b$-axis tension in ns101 are shown in Fig. S3. While the strain induced in ns101 slightly alters the local symmetry of the crystal field, it does not yield measurable shifts in orbital energies.

The central finding of this paper is that the strain state causing large shifts in the energy of the $e_g$ orbitals is the same one that induces the novel superconducting ground state at lower temperatures (we were unable to measure the 100 peak at $T_c$). Our ligand-field picture of the $c$-axis strain preferentially affecting the $d_{x^2-y^2}$ states is consistent with *ab initio* modelling of the band structure of the $t_{2g}$ states under strain [12,13]. In those models shortening the bond length between the ruthenium and equatorial oxygens stabilizes superconductivity by shifting the density of states (DOS) toward the Fermi level, with $d_{xy}$ and $d_{x^2-y^2}$ orbitals raised the most in energy (Fig. 4(b)). This preferential impact of *c*-axis compression on the orbitals of $x^2-y^2$ and $xy$ symmetry in sc110 films can also explain the resemblance of ns101 to bulk. Structural relaxation calculations reveal that the net effect of the strain orientation in ns101 is to stretch the apical oxygens away from the ruthenium without impacting equatorial bond lengths, the main drivers of the energy shift.

Ruf et al. measured the shifting of density of (occupied) states (mostly $d_{xy}$ states) toward the Fermi level under increasing *c*-axis strain in sc110 films using angle-resolved photoemission spectroscopy (ARPES) [12], yet we observe no shifts in the $t_{2g}$ peaks with REXS. In bulk RuO2 with its tetragonally-distorted (apical oxygens compressed toward the ruthenium) octahedral crystal field, the $d_{xy}$ states are the lowest in energy and thus predicted to be fully occupied, shown in Fig. 4(b). Because the REXS energy scan probes empty valence states, it is unsurprising that our data reveal no shift in the lower energy $t_{2g}$ peak (Fig. 3), as the half-filled orbitals populated during the resonance process, $d_{yz}$ and $d_{xz}$, are those least impacted by the strain field. We conclude the strain-dependent shift in the high energy peak of the Ru $L_2$ resonance profile is possibly the empty orbital counterpart to the strain-dependent shift in electronic DOS toward the Fermi energy seen with ARPES that drives superconductivity at low temperature.

A similar strain dependence of the crystal-field splitting has been reported in the electron loss near-edge structure (ELNES) of the Ti $L_3$ edge of SrTiO3 [29]. In that experiment, decreasing the net overlap of oxygen 2*p* orbitals and Ti $d_{x^2-y^2}$ by distortion of the octahedron shifts the $e_g$ peak to lower energy while leaving the $t_{2g}$ unmoved. In the related ferroelectric perovskite BaTiO3 competition between epitaxial strain and polar distortion leads to an atypical crystal field splitting [30]. Similarly, in (110)-oriented RuO2, we argue that two in-plane strains, compression along [001] and tension along [1$\bar{1}$0], compete to produce an unconventional crystal-field splitting that shifts the center of gravity of the $e_g$ manifold to higher energy. This conclusion is further supported by recent work using resonant inelastic x-ray scattering (RIXS) at the Ru *M* edge in bulk $RuO_2$ [31], uncovering the dominant role of the reduced symmetry crystal field in determining the orbital energetics.

In summary we used REXS to study the magnetic Bragg reflection 100 in $RuO_2$ in strain-engineered thin films displaying novel superconductivity at low temperature. We observed modulation of the intensity of the magnetic reflection 100 with azimuth, qualitatively consistent with bulk, but with discrepant periodicities hinting at a strain dependence of the magnetization. Most significantly, we observed a large, ~1 eV shift of the $e_g$ orbitals to higher energy under the same strain state that induces the novel superconductivity. Relaxation of this *c*-axis compression in a thicker, less coherently strained film, produced a smaller shift, confirming the strain dependence of the orbital energies. Our measurements of the unoccupied $e_g$ orbitals show the same trend as *ab initio* modelling of the occupied $t_{2g}$ states that govern the transport properties of these films and further support the use of anisotropic strains to control physical properties in complex materials. Nevertheless, reconciling whether the large magnitudes of these observed shifts can be accounted for solely by strain effects remains to be clarified in future work (both experimental and computational).

We acknowledge helpful conversations with C. A. Occhialini and J. Pelliciari. The work was primarily supported by U.S. Department of Energy, Office of Science, Office of Basic Energy Sciences, under Contract No. DE-SC0019414 (x-ray experiments and interpretation B.G., D.W., Y.S., A.S.; thin film synthesis: H.N., N.J.S). This research used resources of the Advanced Photon Source, a U.S. Department of Energy (DOE) Office of Science User Facility operated for the DOE Office of Science by Argonne National Laboratory under Contract No. DE-AC02-06CH11357. This work was also funded in part by the Gordon and Betty Moore Foundation's EPiQS Initiative, Grant GBMF9073 to Cornell University to support the work of D.G.S, and NSF DMR-22104427 and AFOSR FA9550-21-1-0168 (characterization and model development, J.R., K.M.S.).

**Figures**

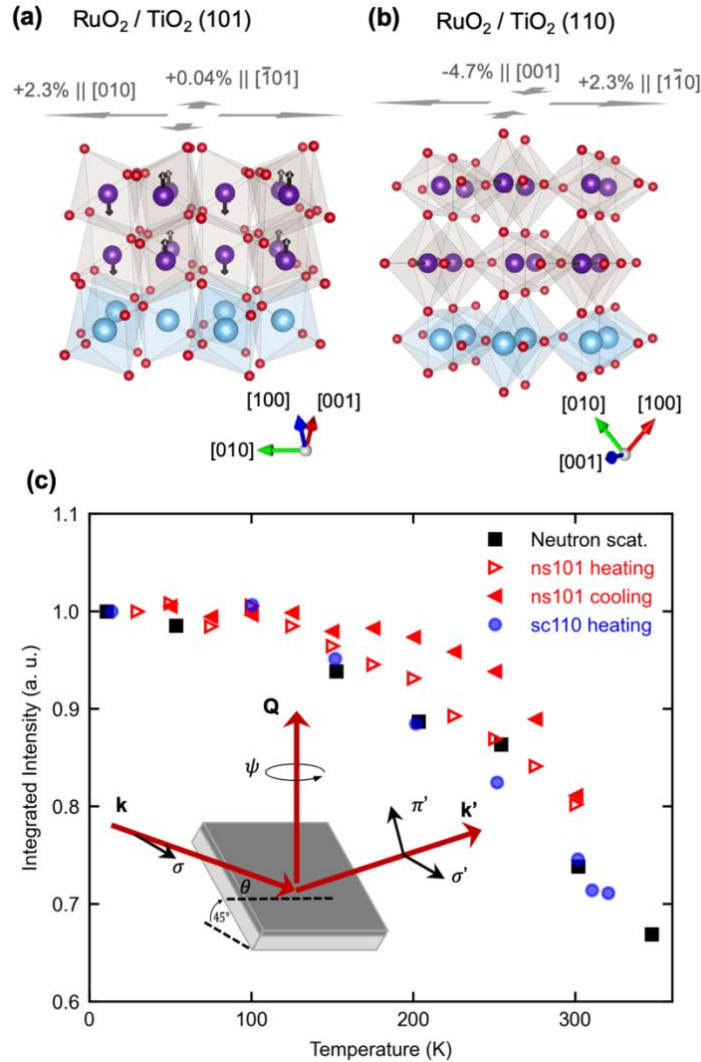

FIG. 1. Resonant magnetic scattering from strained $RuO_2$ thin films. (a) Crystal structure of $RuO_2$ with in-plane strains (gray arrows) synthesized on a (101)-oriented $TiO_2$ substrate. (b) Same as (a) with (110) orientation. The substrate surface normal is oriented toward the top of the page, the magnetic moments (black arrows) of the Ru atoms are aligned along the $c$-axis for clarity. (c) Depicts the temperature dependence of the integrated magnetic 100 Bragg peak. The black squares show the magnetic ordering temperature dependence obtained by neutron scattering, adapted from [15]. The inset illustrates the resonant scattering geometry, where $\theta$ is the Bragg angle, $\mathbf{k}$ is the x-ray wavevector, $\mathbf{Q}$ is the scattering vector, and $\psi$ is the azimuth measured around $\mathbf{Q}$. Photon polarization orthogonal to the scattering plane is denoted $\sigma$, in-plane is denoted $\pi$. The 45° tilt is required for accessing the 100 reflection and severely constrains access to Bragg peak as $\psi$ varies.

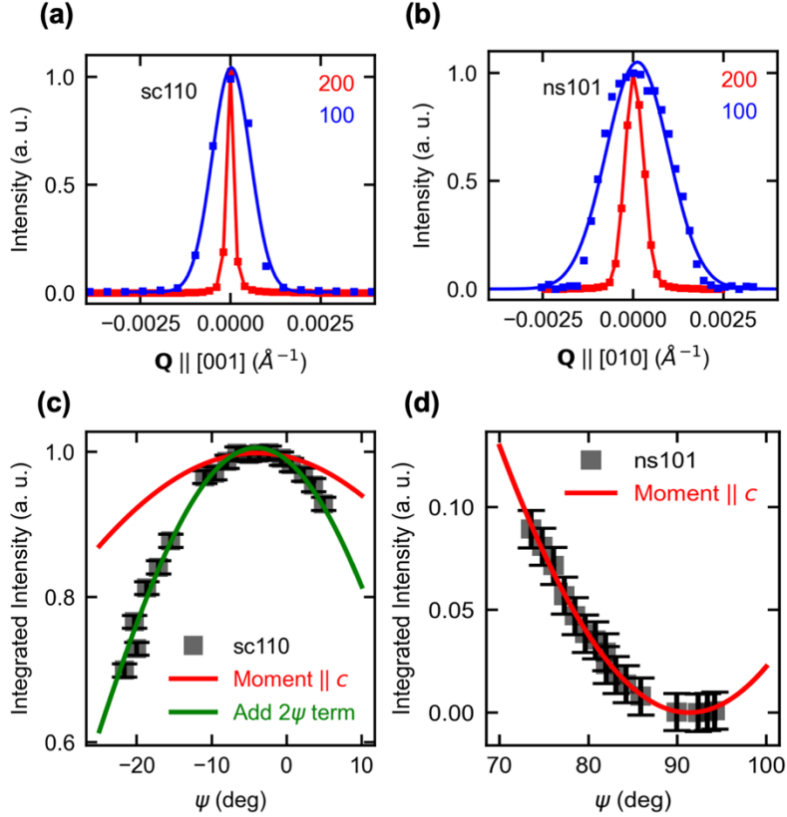

FIG. 2. Antiferromagnetic ordering and azimuthal dependence of the magnetic reflection. (a) and (b) show the structural Bragg peak 200 (red) and the magnetic Bragg peak 100 (blue) of the (110)-oriented (sc110) and (101)-oriented (ns101) samples, respectively. The 100 peaks were measured at 2.98 keV and the 200 peaks at 5.96 keV without readjustment. Solid lines are best Gaussian fits. All peaks are normalized to 1. (c) Shows the integrated intensity of the 100 peak as a function of the azimuth from the sc110 sample. $\psi = 0$ occurs when the crystallographic $c$-axis lies in the scattering plane. The red curve is the theoretical fit for magnetization purely along $c$-axis (Eq. (2) in main text); green includes an additional $\cos(2\psi)$ term (Eq. (3) in main text). (d) Same as (c) for the nc101 sample with intensity rescaled so the red fit achieves a maximum value of 1. Error bars are the $\psi$ averaged, standard deviations of the (110)-oriented data; same errors assumed in (d). Fig. 2(c) shows the average of two repeated measurements over the same angular range and is rescaled to a maximum value of 1. The constant error bars show the average standard deviations in the data over all $\psi$. Fig. 2(d) shows a single scan where we rescale the data so that its extrapolated maximum value at $\psi = 0°$ is 1 when fit with $\cos^2(\psi - \delta)$ in order to assume the same uncertainties measured in sc110.

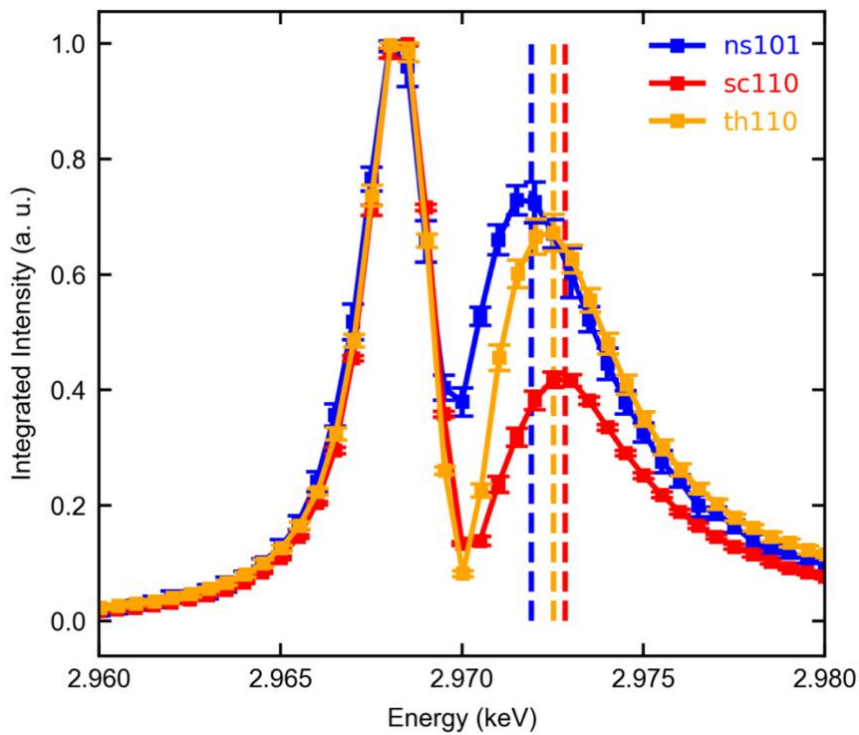

FIG. 3. Strain-tuned resonance profile of the 100 magnetic Bragg peak at the Ru $L_2$ edge. The 0.93 eV shift in the position of the peak at the right is maximized in the highly strained sc110 sample (21.0 nm). The ns101 sample (18.6 nm) is aligned with the bulk. The thicker (48 nm), partially relaxed 110-oriented sample (th110) shows in intermediate peak position and line shape; this resonance was shifted by -0.47 eV to align all $t_{2g}$ peaks. Vertical dashed lines indicate the peak positions determined from Gaussian fits to peaks and are positioned at 2971.91 eV (blue), 2972.52 eV (orange), and 2972.84 eV (red). All scans from film samples are averages over $N$ measurements ($N$=17 for sc110, $N$=28 for th110) obtained at 300 K except ns101 ($N$=12), which is averaged over the range 30-300 K. These resonance profiles have negligible temperature dependence over the range probed (Fig. S1).

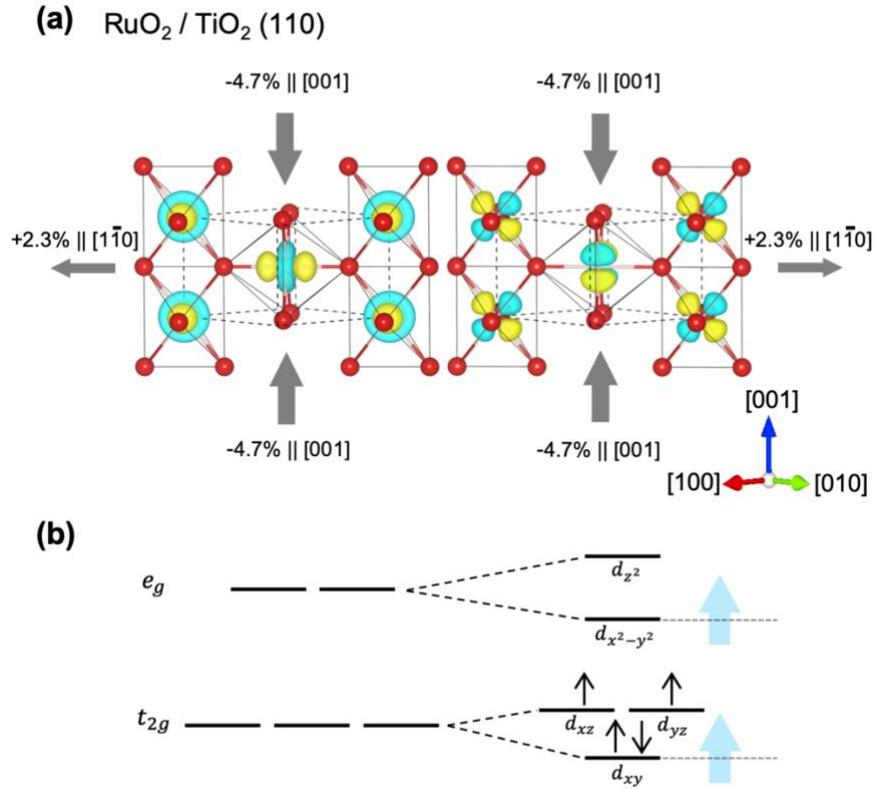

FIG. 4. Schematic depicting how asymmetric epitaxial strain preferentially impacts $4d$ $e_g$ orbitals. (a) Shows the Ru $d_{z^2}$ (left) and $d_{x^2-y^2}$ (right) orbitals within strained oxygen octahedra (red) in the (110)-oriented sample. Gray arrows indicate the strain direction and magnitude. The level splitting in a tetragonally-distorted, octahedral crystal field is shown in (b), with blue arrows indicating the shift in energy due to $c$-axis compression. Though all level degeneracies are lifted in real $RuO_2$, these smaller splittings are suppressed in figure for clarity.

# Supplementary Material:

# Strain-induced orbital energy shift in antiferromagnetic RuO$_2$ revealed by resonant elastic x-ray scattering

Benjamin Gregory[1,2], Jörg Strempfer[3], Daniel Weinstock[2], Jacob Ruf[4], Yifei Sun[2], Hari Nair[2], Nathaniel J. Schreiber[2], Darrell G. Schlom[2,5,6], Kyle M. Shen[1,5], and Andrej Singer[2,*]

[1]*Laboratory of Atomic and Solid State Physics, Department of Physics, Cornell University, Ithaca, NY 14853, USA*
[2]*Department of Materials Science and Engineering, Cornell University, Ithaca, NY 14853, USA*
[3]*Advanced Photon Source, Argonne National Laboratory, Lemont, IL 60439, USA*
[4]*Max-Planck Institute for Chemical Physics of Solids, Nöthnitzer Straße 40, 01187 Dresden, Germany*
[5]*Kavli Institute at Cornell for Nanoscale Science, Cornell University, Ithaca, NY 14853, USA*
[6]*Leibniz-Institut für Kristallzüchtung, Max-Born-Straße 2, 12489 Berlin, Germany*

## Simultaneous fitting of azimuthal scans

To determine the orientation of the magnetic moments under different biaxial strains in RuO$_2$ films, we plot the integrated intensity of the 100 magnetic reflection as a function of azimuthal angle $\psi$ around the scattering vector **Q**. These data are shown in Fig. 2 of the main text and are fitted with Eq. 3. We find that different parameters in Eq. 3 are required to reproduce the azimuthal scans for each strain state. Without canting away from the $c$-axis, the azimuthal scan should reach a maximum at $\psi = 0°$ and a minimum at $\psi = 90°$. We included an offset bounded by 5° in Eq. 3 to account for any misalignment of the sample. Nonetheless, a peak shift can also indicate canting away from the $c$-axis, by the addition of an azimuth-independent, constant term to the sinusoids shifting the minima and changing the periodicity. With this in mind, we attempt to fit both azimuthal scans from both samples simultaneously by the addition of a constant in the following equation,

$$I(\psi) = \left[A_1 \cos(\psi - \delta) + A_2 \cos(2(\psi - \delta)) + C\right]^2,$$
(S1)

and scaling the integrated intensity of the (101)-oriented sample. The best fit we could obtain with Eq. S1 is shown in Fig. S1. The fit parameters are $A_1 = 0.0(3)$, $A_2 = 0.5(1)$ and $\delta = -4.3(9)°$ and $C = 0.5(2)$, with sc110 data normalized to 1.0 at its maximum, and the ns101 data scaled by a factor of 20. The fit is worse for both datasets compared with the fits presented in the main text. The $A_1$ term is eliminated in this fit, presumably because it has the wrong periodicity to fit the sc110 data, and to compensate for what would otherwise be a maximum in the $A_2$ term at $\psi \sim 90°$, the constant offset at $C$ is equal in magnitude to $A_2$. Assuming spin canting is responsible for the azimuthal modulation, these fit parameters would imply the spins are canted 90° away from the $c$-axis in the $a$-$b$ plane. This conclusion is not justified in light of the small azimuthal range accessible to us and the independent measurements that have repeatedly found evidence for a Néel vector predominantly along *c* [1–4]. Because our fitting equation in Eqs. (3) and S1 is not derived from a microscopic model, but is intended to capture terms in the Fourier series that contribute most to the azimuthal dependence, it is more justified to conclude that the failure to fit the two datasets simultaneously implies a difference between the samples resulting from strain.

## Energy scan analysis

We report a 0.93 eV shift in the position of the higher energy peak in the Ru $L_2$ energy scans of the (101)-oriented $RuO_2/TiO_2$ (18.6 nm) film compared with (110)-oriented $RuO_2/TiO_2$ (21.0 nm) film, labelled ns101 and sc110 respectively. These energy scans are shown along with those of a thicker film of (110)-oriented $RuO_2/TiO_2$ (48 nm) labelled th110 in Fig. 3 of the main text. The curves shown in Fig. 3 are averaged energy scans, but in ns101 ($N=12$) the average is over temperature at $\psi = 77°$ and in sc110 ($N=14$) the average is over azimuthal angle at 300 K. In th110 ($N=28$), the average is obtained over different positions on the sample surface at 300 K and $\psi = 0$.

In Fig. S2 we show the unaveraged energy scans of ns101 obtained over different temperatures, the energy scans of sc110 obtained over a range of azimuth and the energy scans of th110 over different sample positions. In all samples the energy scans are highly consistent over these ranges. This confirms that averaging the curves at different temperatures, azimuthal angles, or locations does not affect the position of the resonance peaks.

## Strained crystal field in (101)-oriented $RuO_2/TiO_2$

In Fig. 4(a) of the main text, it is easy to see how the *c*-axis compression reduces the equatorial oxygen ruthenium bond length in sc110. Unfortunately, the strain induced in ns101 is harder to visualize, but we show the $d_{x^2-y^2}$ and $d_{z^2}$ orbitals under +2.3% *b*-axis tension in Fig. S3 for completeness. This strain state alters the local symmetry around the scattering ion, but it does not lead to shifts of orbital energies in the x-ray data. This can be explained by structural relaxation calculations [5], showing that the effect of epitaxial strain in the (101)-oriented sample is only to stretch the apical oxygens, with no change in equatorial bond lengths.

**Figures**

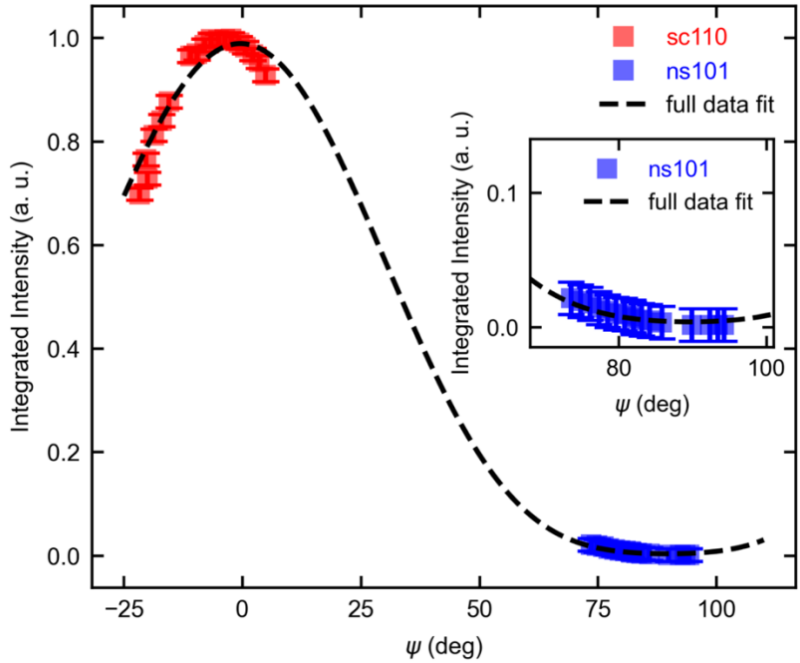

FIG. S1. Simultaneous azimuthal dependence of the (110)- and (101)-oriented RuO$_2$ films. Azimuthal dependence of integral intensity of the 100 magnetic reflection from sc110 and ns110. The ns101 data is scaled by a factor of 20 and sc110 is normalized to 1.0 at its maximum. The dashed line shows the best-fit curve obtained with Eq. S1. The inset shows an enlarged region around $\psi = 90°$.

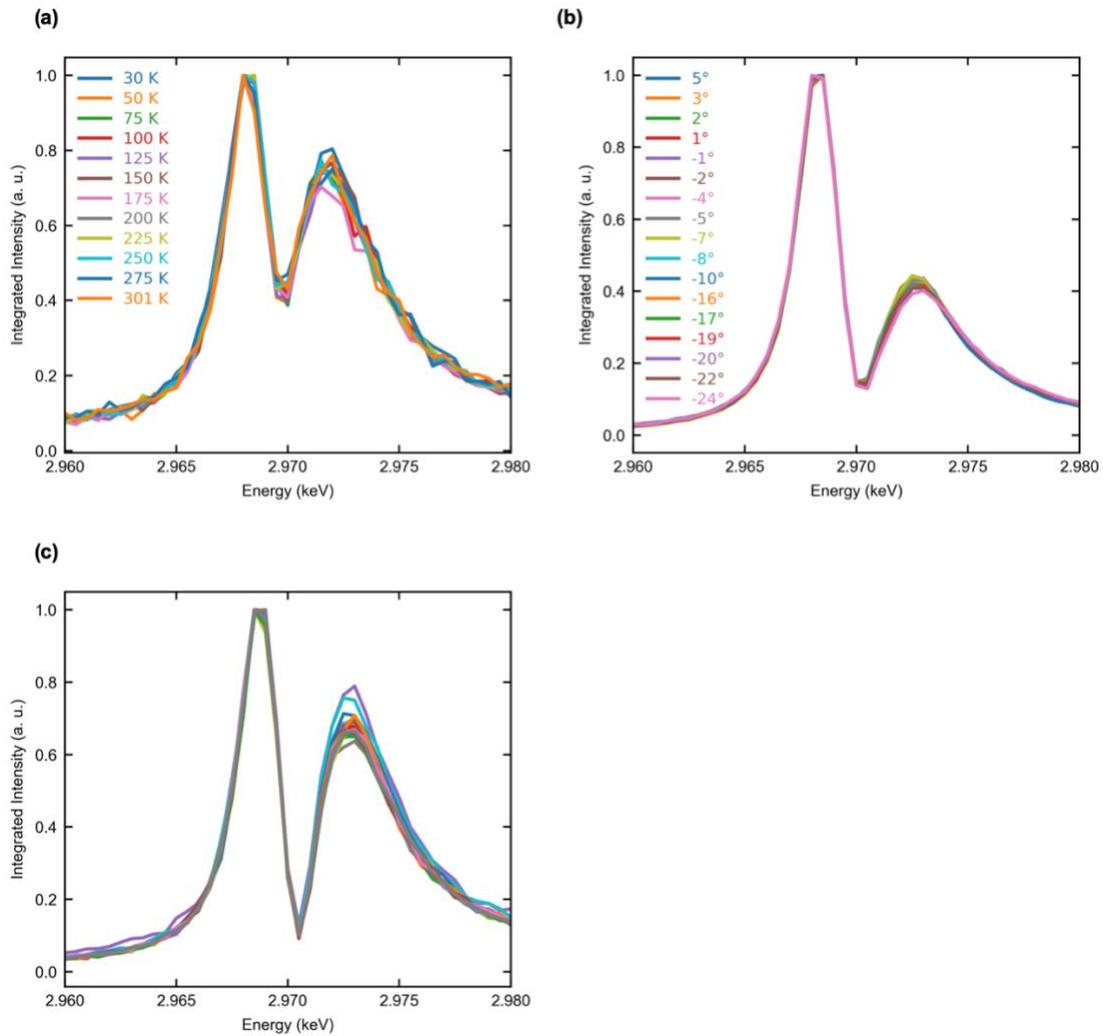

FIG. S2. Azimuthal and temperature dependence of energy scans. (a) Show Ru $L_2$ resonance profiles of the (101)-oriented sample, ns101, obtained over a range of temperatures. (b) Shows Ru $L_2$ resonance profiles of the (110)-oriented sample, sc110, obtained at 300 K at different azimuthal angles $\psi$. (c) Shows Ru $L_2$ resonance profiles of the thicker (110)-oriented sample, th110, obtained at 300 K and $\psi = 77°$ at different sample locations.

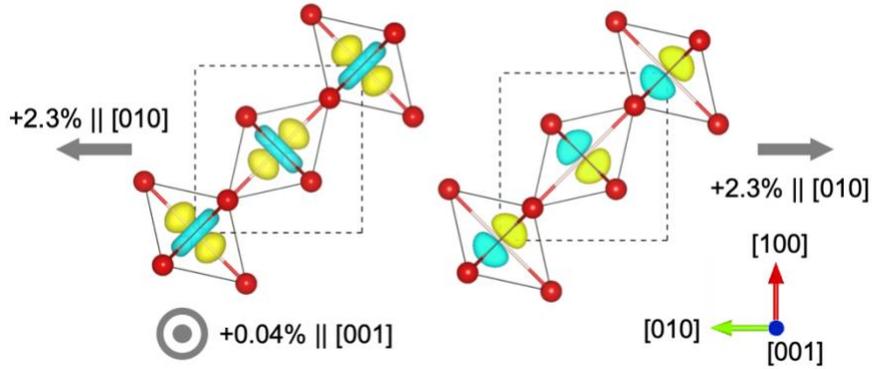

FIG. S3. Asymmetric epitaxial strain on Ru $e_g$ orbitals in (101)-oriented RuO2. (a) Shows the Ru $d_{z^2}$ (left) and $d_{x^2-y^2}$ (right) orbitals within strained oxygen octahedra (red) in the (101)-oriented sample. Gray arrows indicate the strain direction and magnitude.